# Penetration of Non-energetic Hydrogen Atoms into Amorphous Solid Water and their Reaction with Embedded Benzene and Naphthalene

Masashi Tsuge[1*], Akira Kouchi[1], and Naoki Watanabe[1]

[1]Institute of Low Temperature Science, Hokkaido University, Sapporo, Hokkaido 060–0819, Japan

*e-mail: tsuge@lowtem.hokudai.ac.jp

## ABSTRACT

Chemical processes on the surface of icy grains play an important role in the chemical evolution in molecular clouds. In particular, reactions involving non-energetic hydrogen atoms accreted from the gaseous phase have been extensively studied. These reactions are believed to effectively proceed only on the surface of the icy grains; thus, molecules embedded in the ice mantle are not considered to react with hydrogen atoms. Recently, Tsuge et al. (2020) suggested that non-energetic hydrogen atoms can react with CO molecules even in ice mantles via diffusive hydrogenation. This investigation was extended to benzene and naphthalene molecules embedded in amorphous solid water (ASW) in the present study, which revealed that a portion of these molecules could be fully hydrogenated in astrophysical environments. The penetration depths of non-energetic hydrogen atoms into porous and non-porous ASW were determined using benzene molecules to be >50 and ~10 monolayers, respectively (1 monolayer ≈ 0.3 nm).

# 1. Introduction

Aromatic molecules, especially polycyclic aromatic hydrocarbons (PAHs), are ubiquitous in interstellar media (Tielens 2008, 2013). The PAH hypothesis (Leger & Puget 1984; Allamandola et al. 1985, 1989; Puget & Léger 1989) suggests that the unidentified infrared emission features at 3.3, 6.2, 7.7, 8.6, and 11.2 μm observed from several galactic and extragalactic sources are due to UV-excitation-induced infrared fluorescence of PAHs and their derivatives. Benzonitrile ($c$-$C_6H_5CN$), a benzene derivative, was recently detected in the molecular cloud TMC-1 by radio-astronomical means (McGuire et al. 2018). Benzonitrile was successfully detected partially because of its large dipole moment, as apolar or weakly polar nature of PAHs hinders the detection of them by pure rotational transitions. In other words, various types of undetected PAHs and benzene derivatives possibly exist in molecular clouds. For example, the abundance of benzene in the molecular cloud TMC-1 has been estimated to be $5 \times 10^{-10}$ with respect to molecular hydrogen (Jones et al. 2011). In bright photodissociation regions, the abundance of PAHs is estimated from the IR emission features to be $1.4 \times 10^{-5}$ with respect to hydrogen nuclei and the fraction of C locked up in PAHs is ~$3.5 \times 10^{-2}$ (Tielens 2008; Allamandola et al. 1989). Thus, the PAHs account for a significant portion of the interstellar carbon budget and the chemical evolution of them is of particular importance.

In the chemical evolution occurring in molecular clouds, reactions on and within icy grains play an important role in producing a variety of molecules because gaseous-phase reactions, such as ion–molecule reactions, are intrinsically inefficient in producing molecules with large observed abundances (e.g., Watanabe & Kouchi 2008). The chemical species in the molecular clouds are inevitably adsorbed on the icy grains. Therefore, aromatic molecules can also presumably be embedded in icy grains, with certain studies indicating that aromatic molecules can be generated in ice (e.g., Schutte et al. 1993). The molecules embedded or produced in the ice mantle experience energetic processing, such as irradiation with cosmic and UV rays (Briggs et al. 1992; Bernstein et al. 1995; Moore et al. 1996). As the ice mantle of icy grains is predominantly composed of water ice in the amorphous phase (amorphous solid water; ASW), the energetic processing of PAHs embedded in ASW has been extensively studied. However, chemical reactions of PAHs involving non-energetic hydrogen atoms that are accreted onto the icy grain surfaces have not been thoroughly investigated, though the reaction of coronene ($C_{24}H_{12}$) film with hydrogen atoms was reported by Mennella et al. (2012). The reactions of gaseous PAHs and their ions with hydrogen atoms have attracted considerable attention because of their potential role in $H_2$ formation (Cassam-Chenaï et al. 1994; Bauschlicher 1998; Rauls & Hornekær 2008; Le Page et al. 2009; Boschman et al. 2012) and $H_2$ formation involving coronene or superhydrogenated coronene films has also been investigated experimentally (Mennella et al. 2012, 2021).

Tielens & Hagen (1982) proposed that water molecules found in icy mantles are formed in situ through hydrogenation reactions and several water formation mechanisms were investigated experimentally (for a review, see Hama & Watanabe 2013). The importance of the surface hydrogenation reaction was also realized when successive hydrogenation of CO on the surface of ice was found to produce formaldehyde (HCHO) and methanol ($CH_3OH$) (Watanabe et al. 2003, 2004; Fuchs et al. 2009), which have been abundantly detected in astronomical ice (Gibb et al. 2004). The reaction

$$CO \xrightarrow{H} HCO \xrightarrow{H} HCHO \xrightarrow{H} H_3CO/H_2COH \xrightarrow{H} CH_3OH,$$

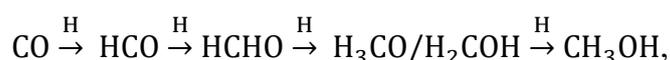

proceeds effectively even at a low temperature of ~10 K. Therefore, this route has been adopted as the most plausible pathway for the formation of the aforementioned organic molecules in chemical models. Subsequently, the hydrogenation reactions of several relevant molecules have been studied. The efficiency of the hydrogenation reaction on the ASW surface decreases at elevated temperatures because the sticking probability of hydrogen atoms decreases and their residence time on the ASW surface is shortened. Therefore, successive hydrogenation of CO molecules on ASW is assumed to occur at temperatures below 20 K, that is, the non-energetic formation of formaldehyde and methanol occurs only at extremely low temperatures. However, Tsuge et al. (2020) demonstrated that the successive hydrogenation occurs even at 70 K when CO molecules are embedded in porous ASW (p-ASW) via "diffusive hydrogenation." Therein, hydrogen atoms were experimentally found to diffuse through cracks in ASW and have a sufficient residence time to react with the embedded CO. The diffusive hydrogenation reaction is important for all atoms and molecules embedded in ASW (Tsuge & Watanabe 2021); therefore, this investigation was extended to benzene ($C_6H_6$) and naphthalene ($C_{10}H_8$) embedded in ASW in the present study to investigate whether diffusive hydrogenation reaction occurs for aromatic species and, based on size dependence, extrapolate to larger PAHs in ASW.

Concrete information regarding the occurrence of diffusive hydrogenation in compact ice (non-porous ASW; np-ASW) is lacking. Laboratory experiments have suggested that the water ice formed in situ during a co-deposition of oxygen molecules and hydrogen atoms is compact (non-porous) because the features attributed to dangling OH bonds do not appear in the infrared (IR) spectrum (Oba et al. 2009). The absence of dangling OH features is consistent with astronomical observations of molecular clouds (Keane et al. 2001). However, the exact morphology of ASW in space, whether totally compact or not, remains unclear. For example, He et al. (2018) demonstrated that the presence of guest molecules can alter and suppress dangling OH features. Tsuge et al. (2020) performed diffusive hydrogenation experiments only on p-ASW because CO molecules mostly sublimated upon the formation of compact ice by annealing at ~100 K. Therefore, in the present study, benzene was used as a probe molecule; an ice sample in which benzene molecules were embedded in np-ASW was prepared, and the occurrence of diffusive hydrogenation in np-ASW was examined.

## 2. Experiments

Experiments were performed using the laboratory setup for surface reactions in interstellar environments (LASSIE); the details of the apparatus have been provided elsewhere (Watanabe et al. 2003; Hidaka et al. 2004; Hama et al. 2014). This system consists of an ultrahigh-vacuum main chamber (base pressure of ~$10^{-8}$ Pa) and a differentially pumped atomic hydrogen source (base pressure of ~$10^{-6}$ Pa). An aluminum (Al) substrate attached to a closed-cycle helium cryostat (RDK-408R, SHI) was placed at the center of the main chamber, and the temperature of the substrate could be varied from 10 K to room temperature. The solid samples on the Al substrate were monitored in situ by reflection–absorption infrared (IR) spectrometry using a Fourier-transform infrared (FTIR) spectrometer (is50R, Thermo Fisher Scientific) equipped with a KBr beam splitter and Hg-Cd-Te detector. Spectra in the region of 650–4000 cm$^{-1}$ were collected with a spectral resolution of 2 or 4 cm$^{-1}$ coadding up to 800 scans.

The solid samples were produced by background vapor deposition, and the liquid $C_6H_6$ and $H_2O$ samples were degassed using several freeze–pump–thaw cycles. The solid $C_{10}H_8$ sample was degassed by vacuum sublimation, in which $C_{10}H_8$ placed in a glass tube was heated for sublimation, and the sublimated $C_{10}H_8$ condensed on the colder parts of the tube. The gaseous components were subsequently pumped out, and the procedure was repeated several times. The column density [X] (molecules cm$^{-2}$) was estimated using the following equation:

$$[X] = \frac{\cos\theta \times \ln 10}{2B} \int A(\tilde{v})d\tilde{v}, \qquad (1)$$

where $\theta$, $B$, and $A(\tilde{v})$ are the incident angles of the IR beam with respect to the substrate (83°), integrated absorption coefficient (cm molecule$^{-1}$), and absorbance at a given wavenumber, respectively. The $B$ value of $C_6H_6$ was obtained from a previously reported value for pure solids (Szczepaniak & Person 1972). Although Hudson & Yarnall (2022) recently reported the absolute intensity of the 1477 cm$^{-1}$ feature of benzene in an $H_2O$-rich environment, pure solid values were used in the present study because the reported intensities are similar to each other within experimental error (~10%), and the absolute intensities of several IR bands were required to estimate the column density as accurately as possible, especially its changes with H-atom irradiation. The $B$ value of $C_{10}H_8$ was adopted from a previously reported value for $C_{10}H_8/H_2O$ (1/15) ice (Sandford et al. 2004). The thicknesses of the deposited samples were evaluated using the IR bands of the ASW. Note that $B$ values reported in the literature were obtained in transmission infrared spectrometry while our experiments used the reflection-absorption infrared spectrometry. This difference would be a source of uncertainty in the estimation of column densities. The number density of one monolayer (ML) was assumed to be $1 \times 10^{15}$ molecules cm$^{-2}$ for both p-ASW and np-ASW. The number density of $H_2O$ molecules in one-ML p-ASW can be smaller than np-ASW due to porosity. Therefore, the thickness of p-ASW, in terms of MLs, tends to be a little underestimated.

Hydrogen atoms were generated from $H_2$ molecules in a microwave-induced plasma in a Pyrex tube and were transferred to the main chamber through a sequence of PTFE and Al tubes. The Al tube was connected to another closed-cycle helium cryostat and was cooled to ~100 K. The fraction of dissociation was ~20%. The flux of the H-atom beam at the substrate surface was estimated to be ~$1 \times 10^{14}$ molecules cm$^{-2}$ s$^{-1}$ using a previously reported method (Hidaka et al. 2007). The H-atom flux was stable over the duration of the experiments (up to 420 min), and the fluctuations during the experiments were within the experimental error of the flux estimation.

Quantum chemical calculations were performed using the Gaussian 16 software package (Frisch et al. 2016). Geometrical optimization and harmonic vibrational analysis were performed using the B3PW91/6-311++G(2d,2p) method (Becke 1993; Perdew et al. 1996). This method has been successfully used to interpret the H-atom addition reaction of PAHs in solid para-hydrogen (e.g., Tsuge et al. 2016). The zero-point vibrational energy was corrected using harmonic wavenumbers, without scaling.

## 3. Results and Discussion

### *3.1 Benzene Embedded in Porous ASW*

Ice samples with $C_6H_6$ embedded in p-ASW were produced by depositing a $C_6H_6/H_2O$ (1/70) gas mixture at a substrate temperature of 20 K. Infrared bands of $C_6H_6$ were observed at 1479, 1038, and 692 (broad) cm$^{-1}$, whereas features due to CH stretching modes were not clearly resolved owing to the intense and broad OH stretching band of p-ASW, which is consistent with the literature (Dawes et al. 2018). After the deposition at 20 K, the ice samples were irradiated with H atoms for 250–420 min (corresponding to an H-atom fluence of (1.5–2.5) × 10$^{18}$ atoms cm$^{-2}$). The resultant IR difference spectrum for a 56-ML-thick sample is shown in Figure 1A, in which the positive and negative bands indicate generation and consumption, respectively. Infrared bands of the cyclohexane ($C_6H_{12}$) product were observed at 2933, 2856, and 1456 cm$^{-1}$, whereas distinct decreases in the $C_6H_6$ content were observed at 3039, 3016, 1479, 1038, and 692 cm$^{-1}$. The weak features observed at 1140, 920, and 877 cm$^{-1}$ were possibly due to cyclohexene ($C_6H_{10}$; Neto et al. 1967), but the existence of small amount of 1,3- and 1,4-cyclohexadiene ($C_6H_8$) (Di Lauro et al. 1968; Stidham 1964) cannot be excluded. The formation of these intermediate species cannot be inferred from their C-H stretching features because C-H stretching modes of CH groups (aromatic C-H stretch) overlap with that of benzene and those of $CH_2$ groups (aliphatic C-H stretch) overlap with cyclohexane. Hama et al. (2014) also observed the efficient formation of $C_6H_{12}$ in the surface hydrogenation of amorphous solid benzene, and used quantum chemical calculations to determine that the first hydrogenation step, $C_6H_6 + H \rightarrow C_6H_7$, which had the highest barrier among the six hydrogen-addition steps, was the rate-limiting step in the sequential hydrogenation of $C_6H_6$ to $C_6H_{12}$.

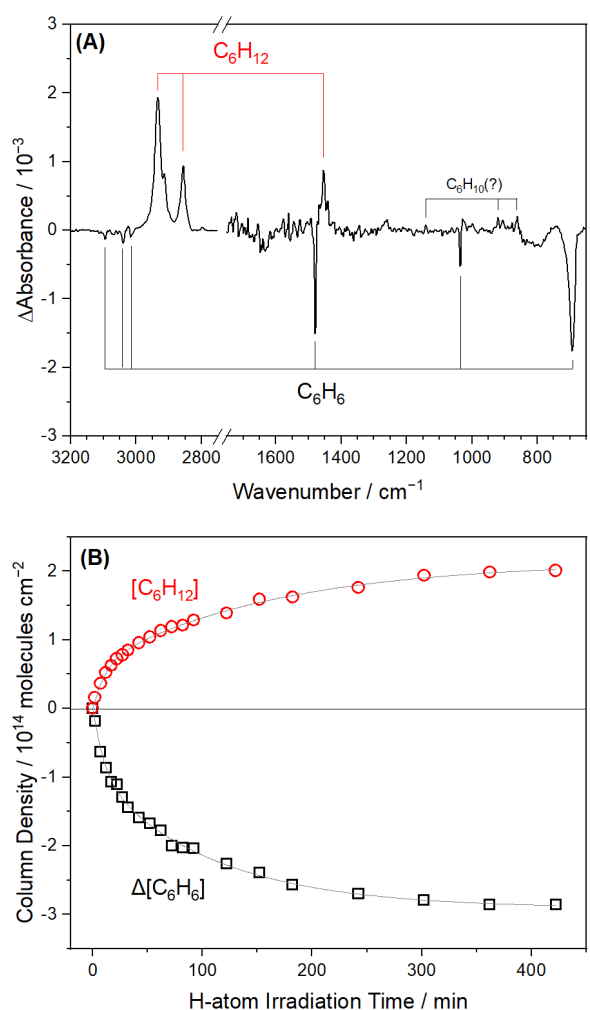

**Figure 1.** (A) Infrared difference spectrum showing the result of 250 min H-atom irradiation of 56-ML-thick $C_6H_6/H_2O$ (1/70) mixed ice. The sample deposition and H-atom irradiation were performed at 20 K. The infrared bands of $C_6H_6$, $C_6H_{10}$ (tentative assignment), and $C_6H_{12}$ are indicated. (B) Variations in column densities (molecules cm$^{-2}$) of $C_6H_6$ and $C_6H_{12}$

upon H-atom irradiation of 56-ML-thick $C_6H_6/H_2O$ (1/70) mixed ice. Open symbols and solid lines represent experimentally derived column densities and fitting results obtained using Eqs. 2 and 3, respectively. An H-atom flux of ~$1 \times 10^{14}$ atoms cm$^{-2}$ s$^{-1}$ was used.

The time evolutions of the column densities of $C_6H_6$ and $C_6H_{12}$ determined using the 56-ML-thick ice irradiated at 20 K are shown in Figure 1B. The column densities were calculated using integrated absorbance and integrated absorption coefficients reported for pure solids: $5.0 \times 10^{-18}$, $1.9 \times 10^{-18}$, and $1.7 \times 10^{-17}$ cm molecule$^{-1}$ for $C_6H_6$ bands at 1479, 1038, and 692 cm$^{-1}$, respectively (Szczepaniak & Person 1972), and $5.3 \times 10^{-17}$, $1.4 \times 10^{-17}$, and $4.2 \times 10^{-18}$ cm molecule$^{-1}$ for $C_6H_{12}$ bands at 2933, 2856, and 1456 cm$^{-1}$, respectively (d'Hendecourt & Allamandola 1986). The decrease in $C_6H_6$ content and generation of $C_6H_{12}$ saturated after 300–400 min of irradiation. The saturated consumption value of $C_6H_6$ in terms of column density was denoted as $\Delta[C_6H_6]_{sat}$, and the column density of $C_6H_{12}$ after saturation was denoted as $[C_6H_{12}]_{sat}$. Based on the Figure 1B data, $\Delta[C_6H_6]_{sat}$ was estimated to be $(2.9 \pm 0.2) \times 10^{14}$ molecules cm$^{-2}$. This value is greater than that of $[C_6H_{12}]_{sat}$ ($(2.0 \pm 0.2) \times 10^{14}$ molecules cm$^{-2}$) because of uncertainties originating from the use of the integrated absorption coefficient of the pure solid in calculating the column densities and because of the partially hydrogenated nature of some of the reacted $C_6H_6$.

The observed temporal variation is similar to that reported for the surface hydrogenation of amorphous solid benzene at 20 K (Hama et al. 2014), indicating that the successive hydrogenation of $C_6H_6$ embedded in p-ASW occurs in a similar manner to that on the surface of amorphous solid benzene. The decrease and increase in the contents of $C_6H_6$ and $C_6H_{12}$, respectively, with time were fitted using exponential functions. Two-phase exponential decay and association functions were used to reproduce the observed changes.

$$\Delta[C_6H_6]_t = A_1 \left(\exp\left(-\frac{t}{\tau_1}\right) - 1\right) + A_2 \left(\exp\left(-\frac{t}{\tau_2}\right) - 1\right), \qquad (2)$$

$$[C_6H_{12}]_t = D_1\left(1 - \exp\left(-\frac{t}{\kappa_1}\right)\right) + D_2\left(1 - \exp\left(-\frac{t}{\kappa_2}\right)\right), \quad (3)$$

where $A_1 + A_2 = [C_6H_6]_0$ and $D_1 + D_2 = [C_6H_{12}]_{sat}$, and $A_1$, $A_2$, $D_1$, and $D_2$ were treated as free fitting parameters. Fitting the data related to the decreasing $C_6H_6$ content yielded $A_1 = (9.5 \pm 0.9) \times 10^{13}$ molecules cm$^{-2}$, $\tau_1 = 11 \pm 2$ min, $A_2 = (1.96 \pm 0.03) \times 10^{14}$ molecules cm$^{-2}$, and $\tau_2 = 107 \pm 9$ min, and that of the increasing $C_6H_{12}$ content yielded $D_1 = (6.2 \pm 0.4) \times 10^{13}$ molecules cm$^{-2}$, $\kappa_1 = 11 \pm 1$ min, $D_2 = (1.51 \pm 0.03) \times 10^{14}$ molecules cm$^{-2}$, and $\kappa_2 = 162 \pm 13$ min; the errors originated from the fitting. It is worth noting that these values corresponded to one experiment and could, therefore, vary with the experimental conditions (e.g., mixing ratio, ice thickness, H-atom flux, and temperature). The temporal variations of 11–56-ML-thick samples were reproduced using two-phase exponents: $\tau_1$ and $\kappa_1$ (~10 min) and $\tau_2$ and $\kappa_2$ (100–300 min). Because no systematic changes were observed for $\tau_2$ and $\kappa_2$ (that is, no correlations between thickness and time constant), the variations in $\tau_2$ and $\kappa_2$ were presumably due to experimental uncertainty. These two components (e.g., $\tau_1$ and $\tau_2$) could be attributed to reactive $C_6H_6$ molecules in two different environments, which will be discussed later.

The $C_6H_6$ consumption ratio for the 56-ML-thick sample ($\Delta[C_6H_6]_{sat}/[C_6H_6]_0$) was estimated to be $0.25 \pm 0.05$ using the ratio of $\Delta[C_6H_6]_{sat}$ to the post-deposition column density of $C_6H_6$ ($[C_6H_6]_0 = 1.1 \times 10^{15}$ molecules cm$^{-2}$ for the 56-ML-thick sample). This high consumption ratio cannot be explained by the hydrogenation reactions occurring near the surface of ice (that is, a few MLs from the ice–vacuum interface); however, diffusive hydrogenation can be invoked in this regard. Therefore, a series of experiments was performed by varying the ice thickness to confirm the occurrence of diffusive hydrogenation. The $C_6H_6$ consumption ratios of $C_6H_6/H_2O$ samples with thicknesses of 11, 20, 28, 37, 46, and 56 MLs were determined (Figure 2; black squares). The $\Delta[C_6H_6]_{sat}/[C_6H_6]_0$ values for the 11-, 20-, 28-, 37-, 46-, and 56-ML-thick samples were $0.24 \pm 0.04$, $0.29 \pm 0.04$, $0.27 \pm$ 0.06, 0.25 ± 0.05, 0.25 ± 0.05, and 0.28 ± 0.05, respectively. These values are evidently thickness-independent, with an average value of 0.26. This result confirms that the diffusive hydrogenation occurred even through the 50-ML-thick p-ASW, which is consistent with the observation of diffusive hydrogenation reactions of CO embedded in 80-ML-thick p-ASW (Tsuge et al. 2020). The normalized saturation value of CO consumption, $\Delta[CO]_{sat}/[CO]_0$, decreases with increasing thickness of $CO/H_2O$ (1/5) mixed ice (Watanabe et al. 2003). At large $CO/H_2O$ ratios, that is, solid CO exposed to vacuum, the possibility of H-atom penetration significantly decreases, presumably because of the smaller sticking coefficient of H atoms and the shorter residence time on the surface. Therefore, diffusive hydrogenation of CO can occur when the $CO/H_2O$ ratio is considerably lower.

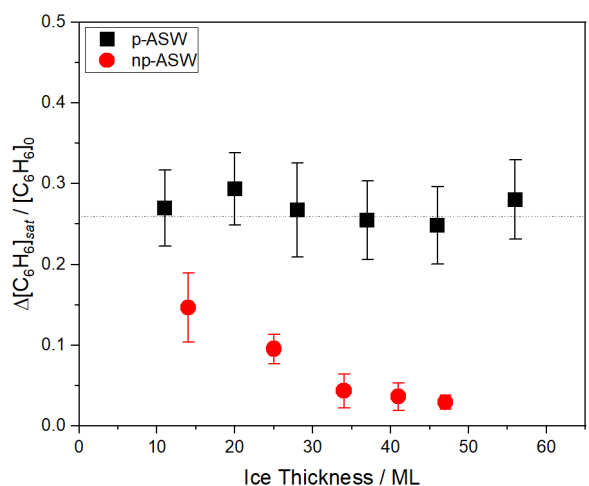

**Figure 2.** Saturation values of $C_6H_6$ consumption normalized to initial column density, $\Delta[C_6H_6]_{sat}/[C_6H_6]_0$, plotted as a function of ice thickness. Black squares and red circles represent the p-ASW- and np-ASW-related results, respectively. The error bars represent errors originating from the column density calculations. The horizontal dotted line indicates the average $\Delta[C_6H_6]_{sat}/[C_6H_6]_0$ value for p-ASW. The np-ASW-related results are described in Section 3.2.

The average fraction of reacted $C_6H_6$ in the $C_6H_6/H_2O$ samples (~0.25) is directly related to the fraction of reactive sites in p-ASW. In the $C_6H_6/H_2O$ samples deposited at 20 K, $C_6H_6$ molecules can exist in an $H_2O$ cage, with each $C_6H_6$ molecule surrounded by $H_2O$ molecules, or on the surface of p-ASW, including the ice–vacuum interface and the surface of cracks or pores. $C_6H_6$ on an Al substrate with submonolayer coverage efficiently reacts with hydrogen, and the reactivity significantly decreases as the coverage approaches a single ML (Hama et al. 2014). The $C_6H_6$–$C_6H_6$ interactions were believed to suppress the reactivity in bulk, and the "dangling" surface $C_6H_6$ that lacks adjacent neighbors remains reactive. In the case of the $C_6H_6/H_2O$ (1/70) samples deposited at 20 K, the $C_6H_6$–$C_6H_6$ interactions could be neglected; consequently, the observed fraction could be related to the distribution of isolation and adsorption sites in p-ASW. The $C_6H_6$ on the surface of p-ASW, including the ice–vacuum interface and the surface of cracks or pores, readily reacted with H atoms similar to the $C_6H_6$ on an Al substrate; the first component exhibited a time constant of ~10 min ($\tau_1$). Experiments conducted using np-ASW (Section 3.2) indicate that the $C_6H_6$ molecules isolated in an $H_2O$ cage reacted with H atoms with a time constant of 100–300 min ($\tau_2$), that is, the slow component. The use of two-phase exponential functions to reproduce the time evolutions of the $C_6H_6$ and $C_6H_{12}$ column densities (Eqs. 2 and 3) suggests that the existence of these two distinct environments was responsible for the origin of the two components in the aforementioned time evolutions. The fast ($\tau_1$ and $\kappa_1$) and slow ($\tau_2$ and $\kappa_2$) components can be attributed to the diffusive hydrogenation of $C_6H_6$ on the surface of cracks or pores and that of $C_6H_6$ molecules isolated in $H_2O$ cages, respectively. Consequently, $A_2/A_1$ represents the ratio of reactive $C_6H_6$ in $H_2O$ cages to that on the surface of cracks or pores. The $A_2/A_1$ ratios corresponding to the 20-, 30-, 37-, 56-, and 58-ML-thick samples were in the range of 1.4–2.4, indicating that the ratio of $C_6H_6$ in $H_2O$ cages to that on the surface was independent of the sample thickness. This result is consistent with the Monte Carlo model of ASW presented by He et al. (2019), who suggested that a fraction of the surface $H_2O$ molecules did not depend on the thickness beyond ~10 MLs. The environments of embedded $C_6H_6$ can, in principle, be deduced from IR spectra, as demonstrated by Tsuge et al. (2020) for the diffusive hydrogenation of CO in ASW. However, the reactive and unreactive isolation sites could not be readily distinguished in the present study; that is, the shape of the $C_6H_6$ IR band, in principle, reflects the interactions between the isolated $C_6H_6$ molecules and the $H_2O$ cage; however, the shape of the $C_6H_6$ bands shown in Figure 1A was similar to that observed after deposition.

### 3.2 Benzene Embedded in Non-porous ASW

$C_6H_6/H_2O$ (1/70) ice samples with an np-ASW structure were prepared by depositing the mixture at 110 K and cooling the deposited samples to 20 K. $H_2O$ deposition at higher temperatures produces np-ASW (Bahr et al. 2008; Bu et al. 2015; He et al. 2019; Nagasawa et al. 2021). The IR spectra collected after deposition showed one feature due to three-coordinated dangling OH bonds at 3696 cm$^{-1}$, which is consistent with the literature; this feature originates from three-coordinated dangling OH bonds located at the top of the ice surface (Nagasawa et al. 2021). H-atom irradiation experiments were performed on 14-, 25-, 34-, 41-, and 53-ML-thick np-ASW samples. The decreasing $C_6H_6$ content and generation of $C_6H_{12}$ were observed to a lesser extent than in the H-atom irradiation experiments with the p-ASW samples.

The $\Delta[C_6H_6]_{sat}/[C_6H_6]_0$ values are plotted as red circles in Figure 2. In contrast to the p-ASW results, the $\Delta[C_6H_6]_{sat}/[C_6H_6]_0$ values decreased with increasing sample thickness (0.15 ± 0.04, 0.10 ± 0.03, 0.043 ± 0.04, 0.036 ± 0.02, and 0.030 ± 0.01 for 14-, 25-, 34-, 41-, and 53-ML-thick samples, respectively). The $\Delta[C_6H_6]_{sat}$ values for these samples were in the range of (2.0–2.5) × 10$^{13}$ molecules cm$^{-2}$, indicating that the depth of H-atom diffusion into np-ASW was less than 14 MLs. Assuming that only $C_6H_6$ in specific sites could be hydrogenated and the fraction of reactive $C_6H_6$ in np-ASW was identical to that in p-ASW (~0.25), the penetration depth of H atoms in np-ASW was estimated to be 6–7 MLs, that is, the $C_6H_6$ density per ML of ASW in the $C_6H_6/H_2O$ (1/70) ice samples was ~$1.4 \times 10^{13}$ molecules cm$^{-2}$, implying that the reactive $C_6H_6$ density per ML of ASW was $3.5 \times 10^{12}$ molecules cm$^{-2}$, which yielded a ratio of 6.4 [(2.0–2.5) $\times 10^{13}$]/[$3.5 \times 10^{12}$]. This value is the lower limit because the fraction of reactive $C_6H_6$ in np-ASW is considered to be smaller than that in p-ASW, as the presence of $C_6H_6$ on the surface of cracks (or pores) was not expected in np-ASW. The lack of cracks or pores in np-ASW indicates that the $C_6H_6$ molecules isolated in $H_2O$ cages reacted with H atoms. Indeed, the time evolution of the $C_6H_6$ molecules, $\Delta[C_6H_6]_t$, embedded in np-ASW could be fitted using a single-exponential function with a time constant of ~100 min, which is similar to the slow component of $\Delta[C_6H_6]_t$ in p-ASW ($\tau_2$; 100–300 min). Based on these considerations, we suggest that the depth of H-atom diffusion into np-ASW was ~10 MLs.

### 3.3 Naphthalene Embedded in Porous ASW

After the deposition of the $C_{10}H_8/H_2O$ (1/60) mixture at 20 K, IR bands of $C_{10}H_8$ were observed at 1599, 1512, 1390, 1271, 1250, 1213, 1130, 1011, and 798 cm$^{-1}$, which were consistent with the literature (Sandford et al. 2004); however, the reported bands at 3071, 3055, and 968 cm$^{-1}$ were not clearly identified because of overlapping with the intense broad features of the ASW. The ice samples were irradiated with H atoms for up to 250 min. Figure 3A shows the IR difference spectrum of ~50-ML-thick ice irradiated for 250 min. The negative bands were due to the decreasing $C_{10}H_8$ content; additional bands at 3072, 3060, and 970 cm$^{-1}$ were also identified. The relatively intense positive bands at 2927, 2860, and 1452 cm$^{-1}$ showed correlated changes upon H-atom irradiation; moreover, these bands and the associated weaker bands (red lines in Figure 3A) were assigned to decahydronaphthalene (decaline, $C_{10}H_{18}$) based on the IR spectrum of $C_{10}H_{18}/H_2O$ ice (see Appendix). Infrared bands of partially hydrogenated naphthalene ($C_{10}H_n$, where $n$ = 9–17) were not detected, indicating that naphthalene was readily hydrogenated once the first hydrogenation step, $C_{10}H_8 + H \rightarrow C_{10}H_9$, occurred.

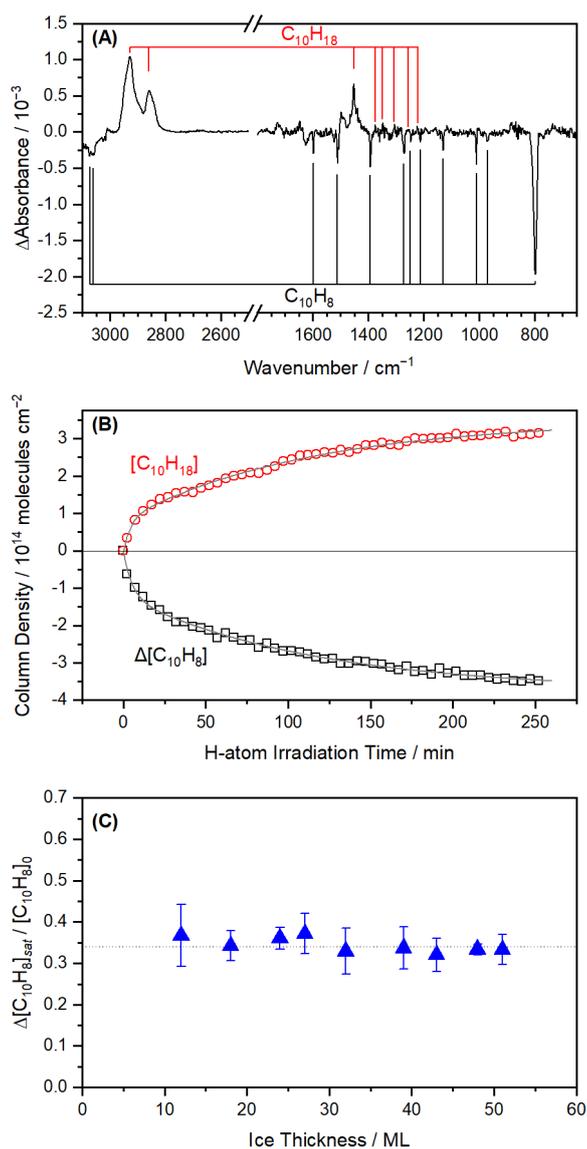

**Figure 3.** Experimental results of H-atom irradiation on naphthalene embedded in p-ASW. (A) Infrared difference spectrum showing the result of 250-min H-atom irradiation of ~50-ML-thick $C_{10}H_8/H_2O$ (1/60) mixed ice. Deposition and H-atom irradiation were performed at 20 K. The infrared bands of $C_{10}H_8$ and $C_{10}H_{18}$ are indicated. (B) Variations in column densities (molecules cm$^{-2}$) of $C_{10}H_8$ and $C_{10}H_{18}$ upon H-atom irradiation of ~50-ML-thick

$C_{10}H_8/H_2O$ (1/60) mixed ice. Open symbols and solid lines represent experimentally derived column densities and fitting results obtained using Eqs. 2 and 3, respectively. An H-atom flux of ~1 × $10^{14}$ atoms cm$^{-2}$ s$^{-1}$ was used. (C) Saturation values of $C_{10}H_8$ consumption normalized to initial column density, $\Delta[C_{10}H_8]_{sat}/[C_{10}H_8]_0$, plotted as a function of ice thickness. The error bars represent errors originating from the column density calculations. The horizontal dotted line indicates the average value of $\Delta[C_{10}H_8]_{sat}/[C_{10}H_8]_0$.

The time evolutions of the column density of $C_{10}H_8$ and IR band intensity of $C_{10}H_{18}$ corresponding to the ~50-ML-thick ice irradiated at 20 K are shown in Figure 3B. The column density of $C_{10}H_8$ was calculated from the IR bands at 1599, 1512, 1390, 1271, 1130, 1011, and 798 cm$^{-1}$, whose integrated absorption coefficients were 5.2 × $10^{-19}$, 1.27 × $10^{-18}$, 8.1 × $10^{-19}$, 4.9 × $10^{-19}$, 4.4 × $10^{-19}$, 5.4 × $10^{-19}$, and 1.09 × $10^{-18}$ cm molecule$^{-1}$, respectively (Sandford et al. 2004). However, the column density of $C_{10}H_{18}$ could not be determined because the integrated absorption coefficients of solid $C_{10}H_{18}$ and $C_{10}H_{18}/H_2O$ ice have not been reported. Instead, the integrated band area from 2970 to 2820 cm$^{-1}$, which was normalized by assuming a one-to-one conversion from $C_{10}H_8$ to $C_{10}H_{18}$, was plotted. Similar to the $C_6H_6$ experiments, two-phase exponential functions were employed to reproduce the observed profiles. The values estimated using the fitting curves obtained using Eqs. 2 and 3 are described henceforth. With respect to the decreasing $C_{10}H_8$ content ($\Delta[C_{10}H_8]$), $A_1$ = (1.22 ± 0.06) × $10^{14}$ molecules cm$^{-2}$, $\tau_1$ = 5.4 ± 0.7 min, $A_2$ = (2.53 ± 0.05) × $10^{14}$ molecules cm$^{-2}$, and $\tau_2$ = 116 ± 8 min; in terms of the increasing $C_{10}H_{18}$ content ([$C_{10}H_{18}$]), $D_1$ = (8.26 ± 0.05) × $10^{13}$ molecules cm$^{-2}$, $\kappa_1$ = 4.3 ± 0.8 min, $D_2$ = (2.67 ± 0.04) × $10^{14}$ molecules cm$^{-2}$, and $\kappa_2$ = 114 ± 6 min; the errors originated from the fitting. Temporal variations observed for sample thicknesses of 24–51 MLs were reproduced using two components: $\tau_1$ and $\kappa_1$ (~5 min) and $\tau_2$ and $\kappa_2$ (~100 min). The reactivities of $C_6H_6$ and $C_{10}H_8$ are compared in the next section.

The $C_{10}H_8$ consumption ratios, $\Delta[C_{10}H_8]_{sat}/[C_{10}H_8]_0$, of the $C_{10}H_8/H_2O$ samples with thicknesses of 12, 18, 24, 27, 32, 39, 43, 48, and 51 MLs were determined (Figure 3C). Because the temporal variation Δ[$C_{10}H_8$] did not saturate after 250 min of irradiation (for thicker samples), Δ[$C_{10}H_8$]$_{sat}$ was estimated using Eq. 2, that is, $A_1 + A_2$ = Δ[$C_{10}H_8$]$_{sat}$. The consumption ratio was independent of ice thickness, with an average value of 0.34 ± 0.02 (Figure 3C; dotted line). This value is greater than that of $C_6H_6$ embedded in p-ASW (0.26). As discussed in Sect. 3.1, $A_2/A_1$ is the ratio of the number of reactive $C_{10}H_8$ in $H_2O$ cages to that on the surface of cracks or pores. The $A_2/A_1$ ratios of samples with thicknesses of 18, 39, 43, 48, and 51 MLs were estimated to be in the range of 1.6–2.5, similar to that for $C_6H_6$ (1.4–2.4), in spite of the different consumption ratios of $C_{10}H_8$ and $C_6H_6$ (0.26 and 0.34, respectively). These results indicate that the $C_{10}H_8$ in $H_2O$ cages and on the surface of cracks or pores was more reactive to H atoms than $C_6H_6$ in the corresponding environment.

### 3.4 Addition of H to PAHs

Most of the reacted benzene and naphthalene molecules were experimentally found to be fully hydrogenated, and the time evolutions of the reactant and product were similar. These results indicate that the first hydrogenation step, $C_6H_6$ + H → $C_6H_7$ or $C_{10}H_8$ + H → $C_{10}H_9$, was the rate-limiting step, as previously suggested for the hydrogenation of pure benzene solid (Hama et al. 2014). The fast component ($\tau_1$) corresponding to the decreasing reactant content can be attributed to the first hydrogenation step for reactants on the surface of the cracks or pores. Moreover, a significant difference was found between the $\tau_1$ time constants for benzene (~10 min) and naphthalene (~5 min). Therefore, quantum chemical calculations were performed for the first hydrogenation step to further investigate this difference.

The barrier height for the first hydrogenation step was determined by calculating the energies of the reactant (e.g., $C_6H_6$ + H) and transition states. In addition to benzene ($C_6H_6$) and naphthalene ($C_{10}H_8$), calculations were performed for pyrene ($C_{16}H_{10}$), coronene ($C_{24}H_{12}$), and ovalene ($C_{32}H_{14}$) to investigate the PAH size dependence of the first hydrogenation barriers. For species other than $C_6H_6$, the hydrogenation process that yielded the most stable isomer was considered. For example, in the case of mono-hydrogenated naphthalene ($C_{10}H_9$), 1-$C_{10}H_9$ (hydrogenation of the carbon atom adjacent to the fused position) has a lower energy than that of 2-$C_{10}H_9$. The calculated results are listed in Table 1. The absolute values varied with the adopted theory; for example, the CAM-B3LYP/6-311++G(d,p)-based calculations performed by Hama et al. (2014) yielded a barrier height of 13.1 kJ mol$^{-1}$ for $C_6H_6$. Therefore, we should compare the relative values calculated with the same method. It is of note that computed values are for gaseous phase molecules and the barrier heights will be affected by the presence of water ice. The barrier height for $C_{10}H_8$ was smaller than that for $C_6H_6$ (14.2 and 18.7 kJ mol$^{-1}$, respectively). At low temperatures, the hydrogenation reaction occurs via quantum mechanical tunneling, the probability of which depends on the width and height of the barrier (Bell 1980). Because the shapes of the potential energy surfaces in the hydrogenation of $C_6H_6$ and $C_{10}H_8$ are similar, the probability of the tunneling reaction correlates with the barrier height. Thus, the computational results suggest that the first hydrogenation of $C_{10}H_8$ is faster than that of $C_6H_6$, which is consistent with the experimentally determined time constants ($\tau_1$) for $C_{10}H_8$ and $C_6H_6$ (~5 and ~10 min, respectively).

**Table 1**

Barrier heights and exothermicity for the first hydrogenation step of benzene ($C_6H_6$), naphthalene ($C_{10}H_8$), pyrene ($C_{16}H_{10}$), coronene ($C_{24}H_{12}$), and ovalene ($C_{32}H_{14}$).

| Species[a] | Barrier height | Exothermicity |
|---|---|---|
| $C_6H_6$ ($C_6H_7$) | 18.7 | 94.8 |
| $C_{10}H_8$ (1-$C_{10}H_9$) | 14.2 | 127.2 |
| $C_{16}H_{10}$ (1-$C_{16}H_{11}$) | 12.2 | 141.2 |
| $C_{24}H_{12}$ (1-$C_{24}H_{13}$) | 15.0 | 117.6 |
| $C_{32}H_{14}$ (7-$C_{32}H_{14}$) | 9.6[b] | 164.1[b] |

**Notes.** All calculations were performed at the B3LYP/6-311++G(2d,2p) level of theory. The zero-point energies were corrected. The energies are expressed in kJ mol$^{-1}$.

[a] Hydrogenation products are specified in parentheses according to the IUPAC nomenclature.

[b] In agreement with the results reported by Tsuge et al. (2016).

A comparison of the calculated barrier heights clearly reveals that an increase in the PAH size decreased the barrier height, except for $C_{24}H_{12}$; moreover, a decent correlation was established between barrier height and exothermicity. A relatively high barrier of $C_{24}H_{12}$ might originate from the extra stabilization resulting from the high symmetry of this molecule. The barrier height for the first hydrogenation step in any PAH is possibly smaller than that of $C_6H_6$. The first hydrogenation step has been found to proceed even at 3.2 K for several PAHs embedded in solid para-hydrogen matrices (Tsuge et al. 2018; Tsuge & Lee 2020). Therefore, any PAH that is embedded in icy grains can possibly be hydrogenated by non-energetic H atoms. A similar suggestion was made by Rauls & Hornekær (2008) and Goumans (2011), who investigated coronene + H and pyrene + H systems, respectively. Further experimental and/or theoretical investigations are required to analyze the degree of hydrogenation (partial or full) in larger PAHs by determining the possibility of the first hydrogenation step being the rate-limiting step (that is, the barrier height for the first step being higher than that for the following steps). In other words, the formation of fully hydrogenated PAHs is expected in this case and partially hydrogenated PAHs are expected when the barrier height rather than the first step is highest. Fourteen possible reaction paths exist, even for the sequential hydrogenation of gaseous $C_6H_6$; moreover, as many as 180 possible paths exist for adsorbed $C_6H_6$ (Saeys et al. 2005). Therefore, calculating all possible sequences of the hydrogenation steps can be difficult. Sebree et al. (2010) investigated sequential addition of four H atoms to $C_{10}H_8$ in the gaseous phase at the G3(MP2,CC)//B3LYP/6-311G(d,p) level of theory. The barrier height for the third hydrogenation step was found to be lower than that for the first hydrogenation step, and the second and fourth steps were barrierless. Nevertheless, experiments analogous to those conducted in the present study can be readily performed on larger PAHs to determine the degree of hydrogenation expected via diffusive hydrogenation.

### *3.5 Astrophysical Implications*

The experimental results of this study suggest that benzene and naphthalene embedded in icy grains can be hydrogenated by the accretion of H atoms on them. Assuming the number density of H atoms to be 1 $cm^{-3}$ in a molecular cloud at 10 K, the H-atom fluence over $10^4$, $10^5$, and $10^6$ years is estimated to be $1.3 \times 10^{16}$, $1.3 \times 10^{16}$, and $1.3 \times 10^{16}$ atoms $cm^{-2}$, respectively. An H-atom flux of $\sim 1 \times 10^{14}$ atoms $cm^{-2}$ $s^{-1}$ was used in the present study, and the irradiation for 2, 20, and 200 min roughly corresponded to the fluence expected in $10^4$, $10^5$, and $10^6$ years, respectively, in the 10 K molecular cloud. Therefore, certain portions of benzene and naphthalene embedded in p-ASW that is as thick as or thicker than 50 MLs can be fully hydrogenated in $10^5$ to $10^6$ years. Hydrogen atoms are also consumed by recombination reactions to form $H_2$ and the effective flux of H-atoms that induce diffusive hydrogenation reaction should be lower. The efficiency of recombination reaction is much higher under the high flux condition in the laboratory as compared to the molecular cloud condition where the H-atom accretion rate is about one H-atom per day to an icy grain. Our experiments demonstrated that benzene and naphthalene molecules can be readily hydrogenated on the surface of cracks or pores in p-ASW environments, whereas a portion of those isolated in $H_2O$ cages can also be hydrogenated. Although the nature of the isolation sites that induce hydrogenation could not be elucidated, molecular dynamics simulations (e.g., Christianson & Garrod 2021) could shed light in this regard.

The experiments performed on benzene in np-ASW showed that the molecules in np-ASW could also be hydrogenated by H atoms accreting on icy grains; however, the H-atom penetration depth was estimated to be ~10 MLs. The significance of this penetration depth in astrophysical environments can be evaluated using the growth rate of icy grains. According to a simulation performed by Furuya et al. (2015), the growth rate of icy grains is ~10 MLs per $10^6$ years. In this case, the limited penetration depth does not affect the fate of the embedded species because ~15% of the benzene molecules embedded in 14 MLs np-ASW are fully hydrogenated within 200 min of irradiation, which corresponds to an H-atom flux over $10^6$ years in a 10 K molecular cloud. However, a considerably higher growth rate of ~100 MLs per $10^5$ years has also been proposed (Jenniskens et al. 1993; Harada et al. 2019). Although the proposed timescale of ice mantle formation ($10^5$ years) is significantly smaller than the typical lifetime of molecular clouds (~$10^7$ years), a chemical model that includes several cycles of accretion and partial or complete evaporation of the ice mantle can reproduce the uniform chemical composition observed in molecular clouds (Harada et al. 2019). In this scenario, a benzene molecule or other PAH molecules accreted on an icy grain are readily buried deep in np-ASW and only partially hydrogenated via diffusive hydrogenation.

The aforementioned considerations indicate that the existence of porous environments facilitates diffusive hydrogenation, and that the reaction is significantly suppressed in non-porous (compact) environments. The ASW found in molecular clouds is generally assumed to be compact owing to the absence of the dangling OH features (Keane et al. 2001).

However, the exact morphology, whether totally compact or not, remains unclear. The apparent surface area of ASW considerably depends on the deposition temperature (Stevenson et al. 1999) and decreases as a function of temperature up to 90 K. In this regard, the deposition temperatures used in the present study (20 and 110 K) can be considered as two extreme cases; the surface areas at these temperatures were estimated using the amount of adsorbed $N_2$ molecules to be >20 MLs and ~1 ML, respectively. Under molecular cloud conditions, the water molecules on icy grains are thought to be formed via several pathways, such as O + H and $O_2$ + H reactions (e.g., Hama & Watanabe 2013). If compaction is induced by the heat of the reaction (Oba et al. 2009), the compactness of the resultant ASW depends on the reaction pathways. Therefore, various factors affect the morphology of ASW in molecular cloud environments, and the ASW need not be assumed to be completely compact. Incorporation of guest molecules is another factor that can affect the morphology of ASW. Small guest molecules, such as $C_6H_6$, can be efficiently packed in small $H_2O$ cages, whereas large PAHs require larger cages for accommodation, which inevitably induces defects near the cage, resulting in a higher possibility for H atoms to attack guest molecules. This could be another reason for the higher reactivity of $C_{10}H_8$ than that of $C_6H_6$. As the size of molecules accreting on the surface of icy grains increases, the molecules require a longer duration to be buried deep into np-ASW (>10 MLs). Therefore, the duration of the diffusive hydrogenation reaction is size-dependent and longer for larger species.

The effects of UV photons are considered in this section. Benzene molecules embedded in ASW decompose to acetylene ($C_2H_2$) and other products or are ionized by UV irradiation. Ruiterkamp et al. (2005) studied the photostability of benzene embedded in p-ASW ($C_6H_6/H_2O \approx 1/15$) and monitored the production of $CO_2$ and CO via oxidation of benzene fragments. Acetylene was not identified in the UV-irradiated $C_6H_6/H_2O$ ice because of its low detection sensitivity and the limited penetration depth of UV light (<0.1 μm). Moreover, the half-life of benzene molecules embedded in ASW in a molecular cloud environment with a typical UV flux of $10^3$ photons cm$^{-2}$ s$^{-1}$ was estimated to be $1.8 \times 10^7$ years. A preliminary UV irradiation experiment on $C_6H_6/C_6H_{12}/H_2O$ (≈1/1/50) ice was performed in the present study using a deuterium lamp (L7293, Hamamatsu Photonics), which indicated that the depletion rates of $C_6H_6$ and $C_6H_{12}$ were similar. Bernstein et al. (2001) performed UV-irradiation experiments on naphthene embedded in ASW ($C_{10}H_8/H_2O$ < 0.01) to simulate the UV fluence expected in dense clouds, and found that only a small percent of embedded naphthalene was photo-processed primarily to naphthol. These results indicate that small aromatic PAHs (including $C_6H_6$) and hydrogenated PAHs embedded in the ice mantle can survive in dense clouds.

In photodissociation regions, strong UV irradiation induces dehydrogenation of gaseous hydrogenated PAHs to yield aromatic PAHs (Tielens 2008, 2013). Analogously, hydrogenated PAHs generated in the ice mantle can also be dehydrogenated by UV irradiation; therefore, diffusive hydrogenation and UV-induced dehydrogenation compete in this regard. For quantitative consideration, we firstly assume that the cross sections for hydrogenation due to accreting H atoms and dehydrogenation due to UV photons are identical, and then consider the validity of this assumption. Because a typical flux of H-atom ($10^4$–$10^5$ atoms cm$^{-2}$ s$^{-1}$) is at most ten times larger than a flux of UV photons ($10^3$–$10^4$ atoms cm$^{-2}$ s$^{-1}$), the assumption above suggests that the timescale of hydrogenation of PAHs embedded in ASW should be considered under an effective H-atom flux of ~10 atoms cm$^{-2}$ s$^{-1}$ or lower. However, the dehydrogenation reaction is suppressed in $H_2O$ matrices owing to the cage effect (e.g., Fillion et al. 2022); that is, the quantum yield of the photodehydrogenation reaction is expected to be lower than that in the gaseous phase because the dissociated H atom can possibly recombine with its counterpart fragment. Additionally, H atoms are produced during the photodissociation of $H_2O$, which can facilitate hydrogenation, as observed in the UV irradiation of $C_6H_{10}/H_2O$ ice (Bernstein et al. 2001). Therefore, an effective H-atom flux (~10 atoms cm$^{-2}$ s$^{-1}$) for the diffusive hydrogenation of PAHs enabled by H atoms and UV-photon fluxes presumably exists as a lower limit and considerably depends on the effective cross section for UV-induced dehydrogenation.

The processing of PAHs in icy grains may be relevant for the organic inventory of solar system bodies, such as asteroids, satellites, and comets (Tielens 2008). Benzene has been detected in comet 67P/Churyumov–Gerasimenko, whereas $C_6H_{12}$ has not (Rubin et al. 2019; Schuhmann et al. 2019). However, these observations cannot be related to the degree of hydrogenation in molecular cloud environments. Although both $C_6H_6$ and $C_6H_{12}$ embedded in the ice mantle could survive the UV field of dense clouds, icy grains experience strong UV irradiation in the later stages (e.g., Ciesla & Sandford 2012), which can lead to considerable decomposition of these species.

This study demonstrated that in modeling the chemical evolution of PAHs embedded in icy grains, the hydrogenation reactions induced by H atoms accreting on the icy grains is worth considering in addition to the processes induced by UV and cosmic ray irradiation (Gudipati & Allamandola 2003, 2004; Bouwman et al. 2010). Consequently, the PAHs ionized by UV and cosmic ray irradiation can react with non-energetic H atoms, or the neutral PAHs undergo partial hydrogenation, followed by ionization. For example, the diffusive hydrogenation of PAH cations embedded in ASW must be investigated by electronic spectroscopy (e.g., Gudipati & Allamandola 2003; Garusha et al. 2013) to mitigate the difficulties involved in distinguishing the charge state (cation or neutral) from the IR spectra.

The aforementioned reaction sequence is important when considering the role of PAH in $H_2$ formation. The formation of $H_2$ in interstellar media occurs efficiently on the surface of dust grains via the H + H recombination reaction (e.g., Wakelam et al. 2017) because the gaseous-phase process is too slow to account for the large abundance. Additionally, PAHs and their derivatives are believed to catalyze $H_2$ formation. Several plausible processes for $H_2$ formation, including hydrogenated neutral PAHs, PAH cations, and protonated PAHs, have been proposed based on theoretical and experimental investigations (Cassam-Chenaï et al. 1994; Bauschlicher 1998; Rauls & Hornekær 2008; Le Page et al. 2009; Fu et al. 2012; Szczepanski et al. 2011; Boschman et al. 2012; Mennella et al. 2012, 2021). For instance, Mennella et al. (2012, 2021) reported $H_2$ production via H(D)-atom irradiation of a coronene ($C_{24}H_{12}$) film and UV irradiation of a fully hydrogenated coronene ($C_{24}H_{36}$) film. These results indicate that both partially and fully hydrogenated PAHs contribute to $H_2$ formation. However, to quantify the contribution of PAHs to overall $H_2$ formation, further experimental and theoretical investigations are required to accurately simulate the evolution of the hydrogenation state of PAHs.

This study was supported by the Japan Society for the Promotion of Science (JSPS KAKENHI; Grant Nos. JP21H01139, JP18K03717, and JP17H06087).

# Appendix

Infrared spectra of solid decaline ($C_{10}H_{18}$) and $C_{10}H_{18}/H_2O$ ice were measured to assign the product of H-atom irradiated $C_{10}H_8/H_2O$ ice (Figure A1). The mixing ratio of the $C_{10}H_{18}/H_2O$ ice is approximately 1/50, which is estimated from partial pressures of premixed gaseous sample.

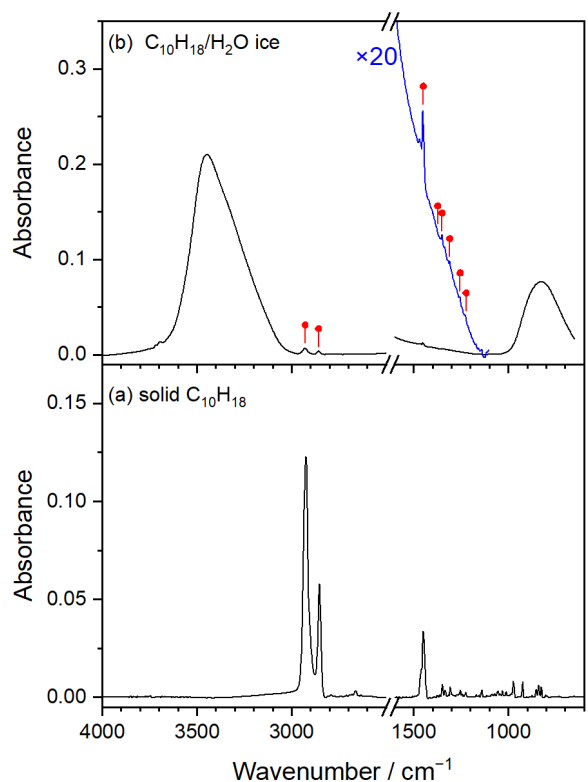

**Figure A1.** Infrared spectra of (a) solid $C_{10}H_{18}$ and (b) $C_{10}H_{18}/H_2O$ (1/50) ice. These samples were deposited at 20 K. In (b), the IR features observed in H-atom irradiated $C_{10}H_8/H_2O$ ice were marked with filled circles.